\documentclass[preprint,amsmath,superscriptaddress,amssymb,floatfix,nofootinbib]{revtex4}
\usepackage{epsf}
\usepackage{graphicx}
\usepackage{subfigure}
\usepackage{feynmf}
\usepackage{tikz}
\usepackage{epstopdf}
\newcommand{\beq}{\begin{equation}}
\newcommand{\eeq}{\end{equation}}

\begin{document}

%\preprint{~~PITT-PACC-????}

\title{$D^{\pm}$ Production Asymmetry at the LHC from Heavy-Quark Recombination}

\def\pitt{Pittsburgh Particle Physics Astrophysics and Cosmology Center (PITT PACC)\\
Department of Physics and Astronomy, University of Pittsburgh, Pittsburgh, PA 15260, USA
\vspace*{.2cm}}
\def\wsu{Department of Physics and Astronomy, \\
Wayne State University, Detroit, MI 48201, USA
\vspace*{.2cm}}
\def\mctp{Michigan Center for Theoretical Physics, \\
University of Michigan, Ann Arbor, MI 48109, USA
\vspace*{.2cm}}

\author{W. K. Lai\footnote{Electronic address: wal16@pitt.edu}}
\affiliation{\pitt}

\author{A. K. Leibovich\footnote{Electronic address: akl2@pitt.edu}}
\affiliation{\pitt}

\author{A. A. Petrov\footnote{Electronic address: apetrov@wayne.edu}}
\affiliation{\wsu}\affiliation{\mctp}

\date{\today}

\begin{abstract}
The asymmetry in the forward region production cross section of $D^{\pm}$ 
is calculated using the heavy-quark recombination mechanism for $pp$ collisions at $7$~TeV. 
By suitable choices of four non-perturbative parameters, our calculated results can reproduce
those obtained at LHCb.
We find $A_p\sim-1\%$ when integrated over $2.0\textrm{ GeV}<p_T<18\textrm{ GeV}$ and $2.2<\eta<4.75$,
which agrees with $A_p=-0.96\pm0.26\pm0.18\%$ as measured by LHCb. Furthermore, the calculated distributions
in $\eta$ and $p_T$ agree reasonably well with those obtained at LHCb. 
\end{abstract}

\maketitle

%==============================================================
% contents
%==============================================================

Observation and proper interpretation of CP-violation in charm system could provide an outstanding opportunity for indirect searches for physics beyond the standard model. Even already available bounds on CP-violating interactions provide rather stringent constraints on the models of new physics because of availability of large statistical samples of charm data from LHCb, Belle, and BaBar experiments.  Larger samples will be available soon from both $pp$ and $e^+e^-$ machines~\cite{Butler:2013kdw}. 

One of the simplest signals for CP-violation in charm is obtained by comparing partial decay widths of charm mesons to those of anti-charm mesons. While CPT-symmetry requires the total widths of $D$ and $\overline D$ to be the same, the partial decay widths $\Gamma(D \to f)$ and $\Gamma(\overline D \to \overline f)$ are different in the presence of CP-violation, which is signaled by a non-zero value of the asymmetry
\begin{equation}
a^f_{CP} = \frac{\Gamma(D \to f)-\Gamma(\overline D \to \overline f)}{\Gamma(D \to f)+\Gamma(\overline D \to \overline f)}.
\end{equation}
This signal is reasonably robust for $D^+/D^-$ mesons, provided that the number of decaying particles and anti-particles is the same. However, at the Large Hadron Collider (LHC), the number of produced $D^+$ and $D^-$ mesons might not be the same due to the fact that the initial state contains two protons. With CP-violating asymmetries expected to be at the per mille levels~\cite{Artuso:2008vf}, it is important to examine production asymmetry of $D$-mesons both experimentally and theoretically.

Indeed, fixed-target experiments have already observed large asymmetries of charmed mesons and baryons in the forward region. In hadroproduction, the charmed hadrons are preferentially produced with a light valence quark of the same type as what appears in the hadronic beam, for example \cite{Aitala:1996hf}. This has been termed the {\it leading particle effect}. More recently, a similar asymmetry in $D^\pm$ production, defined as
\begin{equation}
A_p=\frac{\sigma(D^+)-\sigma(D^-)}{\sigma(D^+)+\sigma(D^-)}\,,  \label{eq:asymmetry}
\end{equation}
has been measured in the forward region to be $\sim -1\%$ by the LHCb collaboration~\cite{LHCb:2012fb}.  What are the theoretical expectations for this asymmetry?

Factorization theorems of perturbative QCD~\cite{Collins:1985gm} state that heavy hadron production cross section can be written in a factorized form. At the LHC, the cross section for producing a $D$ ($c\bar{q}$) meson in a $pp$ collision, at leading order in a $1/p_T$ expansion, is given by
\begin{equation}
d\sigma[pp\rightarrow D+X]=\sum\limits_{i,j}f_{i/p}\otimes f_{j/p}\otimes d\hat{\sigma}[ij
\rightarrow c\bar{c}+X]\otimes D_{c\rightarrow D}\,,    \label{eq:pQCD}
\end{equation}
where $f_{i/p}$ is the partion distribution function for parton $i$ in the proton, $d\hat{\sigma}
(ij\rightarrow c\bar{c}+X)$ is the partonic cross section and $D_{c\rightarrow D}$ is the 
fragmentation function describing hadronization of a $c$ quark into a $D$ meson. The corresponding equation for 
$\bar{D}$ is obtained by replacing $D_{c\rightarrow D}$ by $D_{\bar{c}\rightarrow \bar{D}}$. Charge conjugation $C$ is expected to 
be a good symmetry in QCD, so $D_{c\rightarrow D} = D_{\bar{c}\rightarrow \bar{D}}$. Thus, perturbative QCD predicts 
that $A_p=0$, which is at least true at leading order in $1/p_T$ expansion.

This conclusion led theorists to examine other mechanisms for generating production asymmetry of Eq.~(\ref{eq:asymmetry}), including 
attempts to describe the effect phenomenologically. The main idea of those approaches is to identify phenomenological mechanisms 
that can lead to enhanced production asymmetries, such as ``meson cloud" effects. The results of these model-dependent calculations 
can be found in Refs.~\cite{Norrbin:2000zc,Cazaroto:2013wy}. We note that it might be challenging to interpret some of those mechanisms 
in QCD.

To reconcile the experimental observations with QCD, we note that there are corrections to Eq.~(\ref{eq:pQCD}) that scale as powers of 
$\Lambda_{\rm QCD}/m_c$ and $\Lambda_{\rm QCD}/p_T$. In principle, one can expect non-vanishing power-suppressed contributions to 
$A_p$ at low $p_T$.  A QCD-based model for these power corrections is the heavy-quark recombination 
mechanism~\cite{Braaten:2001bf,Braaten:2001uu,Braaten:2002yt,Braaten:2003vy}.  In this scenario, a light quark involved in the hard 
scattering process combines with the heavy quark produced in that interaction to form the final state meson, leading to corrections of order 
$\Lambda_{\rm QCD} m_c/p_T^2$. In what follows, after a quick review of the heavy-quark recombination mechanism,\footnote{For a full 
review, please see Refs.~\cite{Braaten:2001bf,Braaten:2001uu,Braaten:2002yt,Braaten:2003vy}.} we calculate $A_p$
due to heavy quark recombination. 

\begin{figure}
\begin{center}
\subfigure[]{
\includegraphics[scale=0.27]{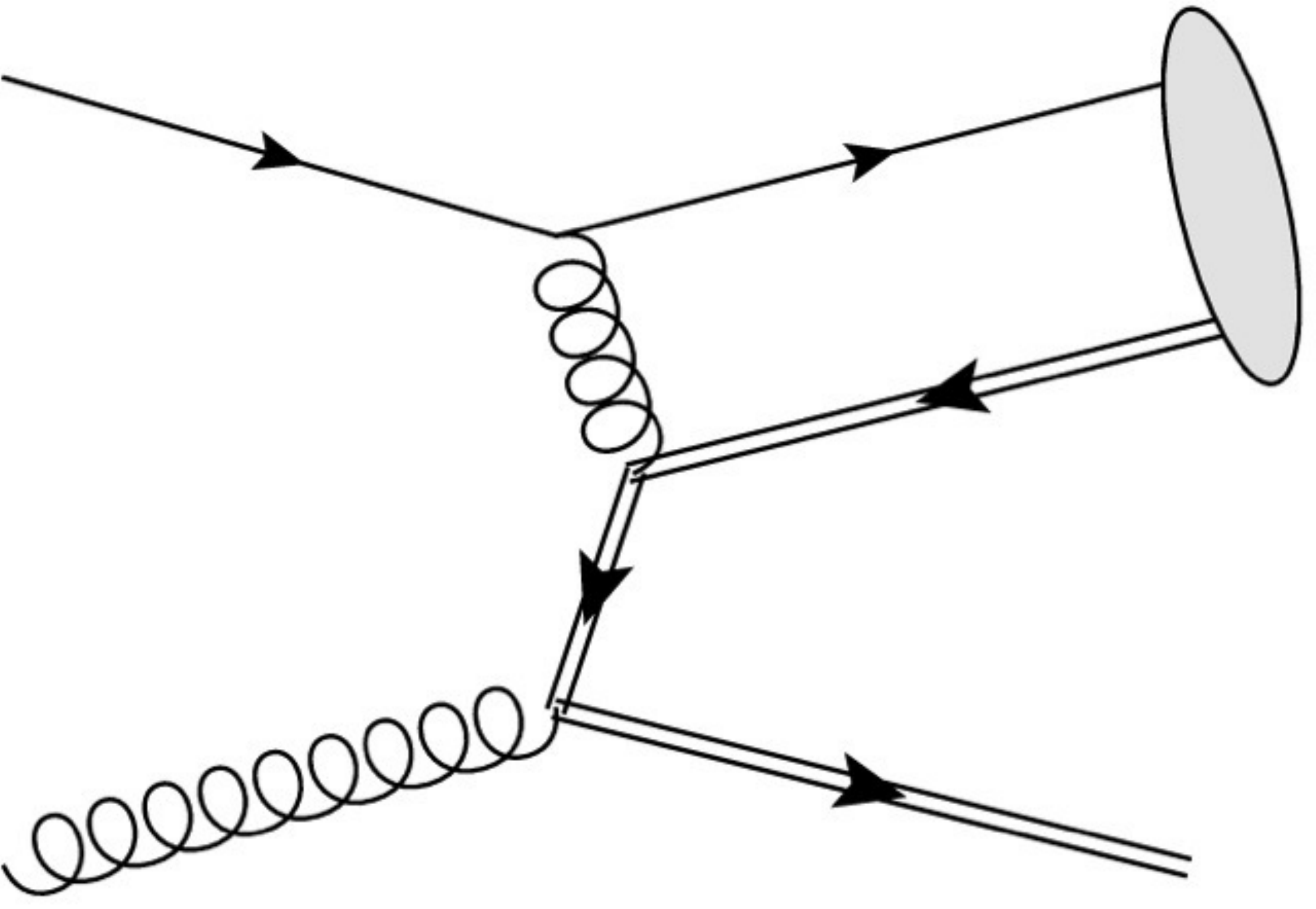}}\label{fg:recomb_qg}
\hskip .5in
\subfigure[]{
\includegraphics[scale=0.05]{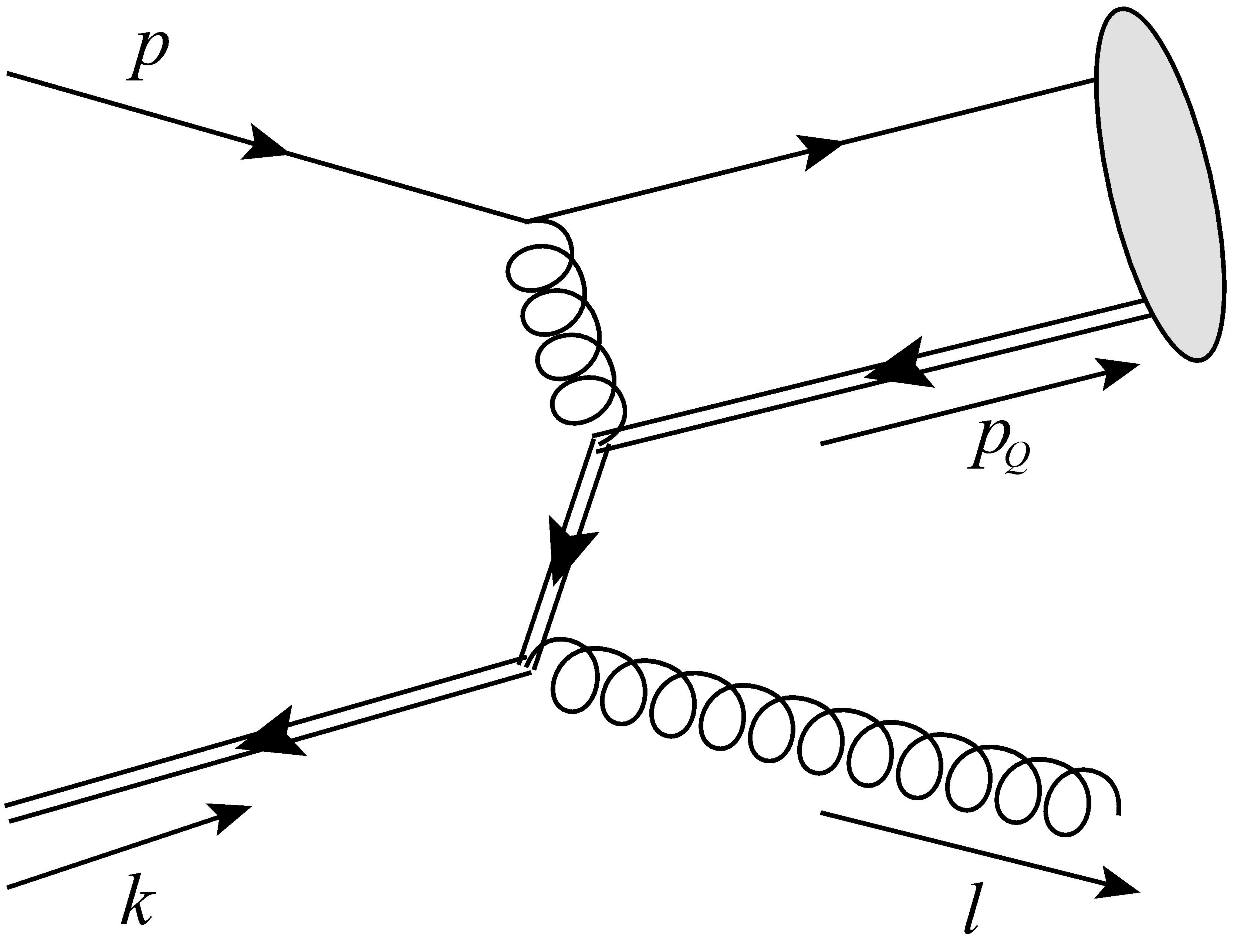}}\label{fg:recomb_Qq}
\vskip .05in  
\caption[]{Diagrams for production of a $\bar{D}$ meson by the heavy-quark recombination mechanism for (a) $qg\rightarrow (\bar c q)^n+c$
and (b) $q\bar{c}\rightarrow (\bar c q)^n+g$. Each process has five diagrams. Single lines represent light quarks, 
double lines heavy quarks, and the shaded blob the $\bar{D}$ meson. }
\label{fg:recomb}
\end{center}
\end{figure}

Imagine production of a heavy meson with the light quark of the same flavor as that appears in the beam.  For instance, for a proton beam 
we could have $D^-$ or $\overline D^0$ states, which we shall generically call $\overline D$. The recombination process, shown in Fig.~\ref{fg:recomb}~(a), comes in as a power-suppressed correction to Eq.~(\ref{eq:pQCD}). As mentioned, the light quark in the production of $\overline D$ 
comes from the incident proton.  The contribution to the cross section is given by
\beq
d\hat\sigma[\bar D]=d\hat{\sigma}[qg\rightarrow (\bar c q)^n+c]\rho[(\bar cq)^n\rightarrow \bar D]\,, \label{eq:recomb}
\eeq
where $(\bar cq)^n$ indicates that the light quark of flavor $q$ with momentum $\Lambda_{\rm QCD}$ in the $\bar c$ rest frame is produced in the states $n$, where $n$ labels the color and angular momentum quantum numbers of the quark pair. The cross section is factored into  a 
perturbatively calculable piece $d\hat{\sigma}[qg\rightarrow (\bar c q)^n+c]$ and the nonperturbative factor $\rho[(\bar cq)^n\rightarrow \bar D]$
encoding the probability for the quark pair with quantum number $n$ to hadronize into the final state  including the $\overline D$.  
The perturbative piece was calculated to lowest order in~\cite{Braaten:2001bf}.
Equation~(\ref{eq:recomb}) must then be convoluted with the proton parton distribution functions to get the final hadronic cross section.

Besides the $qg\rightarrow (\bar c q)^n+c$ process, there are also contributions from $q\bar{c}\rightarrow (\bar c q)^n+g$, as shown in Fig.~\ref{fg:recomb}~(b). Using the method introduced in \cite{Braaten:2001bf}, the partonic cross sections from initial state charm are calculated to be
\begin{eqnarray}
\frac{d\hat\sigma}{d\hat t}[\bar Q q(^1S_0^{(1)})] &=& \frac{2\pi^2\alpha_s^3}{243}\frac{m_Q^2}{S^3}\left[\frac{64 S^2}{T^2} - \frac{m_Q^2 S}{UT} \left(79 - \frac{112 S}{U} - \frac{64 S^2}{U^2}\right)+ \frac{16 m_Q^4}{U^2}\left(1 - \frac{8S}{U}\right)
\right],\nonumber\\
\frac{d\hat\sigma}{d\hat t}[\bar Q q(^3S_1^{(1)})] &=& \frac{2\pi^2\alpha_s^3}{243}\frac{m_Q^2}{S^3}\left[\frac{64 S^2}{T^2}\left(1 + \frac{2 S^2}{U^2} \right)
- \frac{m_Q^2}{T} \left(28 - \frac{4 U}{S} - \frac{19 S}{U} -\frac{368 S^2}{U^2} + \frac{64 S^3}{U^3}\right)\right.\nonumber\\
&&\qquad\qquad\qquad\left.
+ \frac{48 m_Q^4}{U^2}\left(1 - \frac{8S}{U}\right)\right],\nonumber\\
\frac{d\hat\sigma}{d\hat t}[\bar Q q(^1S_0^{(8)})] &=& \frac{4\pi^2\alpha_s^3}{243}\frac{m_Q^2}{S^3}\left[\left(9 + \frac{9 S}{T} + \frac{4S^2}{T^2}\right) - \frac{m_Q^2}{T} \left(\frac{9U}S - \frac{79 S}{2U} - \frac{7 S^2}{U^2}-\frac{4S^3}{U^3}\right)\right.\nonumber\\
&&\qquad\qquad\qquad\left.
- \frac{m_Q^4}{U^2}\left(8 + \frac{8S}{U}+ \frac{9U}S \right)
\right],\nonumber\\
\frac{d\hat\sigma}{d\hat t}[\bar Q q(^3S_1^{(8)})] &=& \frac{4\pi^2\alpha_s^3}{243}\frac{m_Q^2}{S^3}\left[\left(16 + \frac{13 U}{T} +\frac{14 T}{U} + \frac{12U^2}{T^2}+ \frac{8T^2}{U^2}\right) \right.\nonumber\\
&&\qquad\qquad\qquad\left.
+ \frac{m_Q^2}{T} \left(158 + \frac{133U}{S} + \frac{233 S}{2U} + \frac{5 S^2}{U^2}-\frac{4S^3}{U^3}\right)\right.\nonumber\\
&&\qquad\qquad\qquad\left.
- \frac{3 m_Q^4}{U^2}\left(8 + \frac{8S}{U}+ \frac{9U}S \right)
\right],
\end{eqnarray}
where we have defined $S = \hat s - m_Q^2 = (k+p)^2 - m_Q^2, T = \hat t = (k - p_Q)^2,$ and $U = \hat u - m_Q^2 = (k - l)^2 - m_Q^2$.

%==========================================================================

%============================================================================

The $c$ quark in Eq.~(\ref{eq:recomb}) could fragment into a $D$ meson, this time of opposite flavor, i.e., a $D^+$ or $D^0$, generically labeled $D$.    Thus, to get the full rate due to recombination for producing $\overline D$ mesons, we also need to account for the contribution where a light antiquark comes from the proton, while the $\bar c$ fragments into a $\overline{D}$. We thus have three contributions,
\begin{subequations}
\begin{align}
a)\phantom{aaaa}d\hat\sigma[\bar D]&=d\hat{\sigma}[qg\rightarrow (\bar c q)^n+c]\rho[(\bar c q)^n\rightarrow \bar D]\,,\label{eq:recombination} \\
b)\phantom{aaaa}d\hat\sigma[\bar D]&=d\hat{\sigma}[q\bar{c}\rightarrow (\bar c q)^n+g]\rho[(\bar c q)^n\rightarrow \bar D]\,,
\label{eq:recombination_Qq} \\
c)\phantom{aaaa}d\hat\sigma[\bar D]&=d\hat{\sigma}[\bar qg\rightarrow (c\bar q)^n+\bar{c}]\rho[(c\bar q)^n\rightarrow H]\otimes
D_{\bar c\rightarrow \bar D}\,,
\label{eq:recombinationfragment}
\end{align}
\end{subequations}
where $H$ can be any hadron. The recombination cross section for producing a $D$ is obtained by taking the charge conjugate of the above equations.
Below, we will neglect $C$-violation and take $\rho[(\bar{c}q)^n\rightarrow \bar{D}]=\rho[(c\bar{q})^n\rightarrow D]$. For simplicity,
in process c) we will restrict $H$ to be $D$ only and sum over $\bar{q}=\bar{u},\bar{d}$ and $\bar{s}$ with $SU(3)$ flavor symmetry assumed.   

As discussed in \cite{Braaten:2002yt}, the nonperturbative parameters $\rho[(\bar c q)^n\rightarrow \bar D]$ with the same flavor and angular momentum quantum numbers as the $\bar D$ scale as $\Lambda_{\rm QCD}/m_c$.  However, the amplitudes for $(\bar c q)^n$ production with $L>0$ are suppressed relative to the S-wave states. On the other hand, $^3S_1 \to D$ transition is achieved via emission of magnetic-type gluons, which,
contrary to the heavy quarkonia case,  is not suppressed for $D$-mesons. Thus, the leading contributions to productions of $D^\pm$ mesons by heavy-quark recombination consists of four possible options of $n$:
\begin{align}
\rho^{sm}_1&=\rho[c\bar{d}(^1S_0^{(1)})\rightarrow D^+],
&\rho^{sf}_1&=\rho[c\bar{d}(^3S_1^{(1)})\rightarrow D^+], \nonumber \\
\rho^{sm}_8&=\rho[c\bar{d}(^1S_0^{(8)})\rightarrow D^+],  
&\rho^{sf}_8&=\rho[c\bar{d}(^3S_1^{(8)})\rightarrow D^+]. \label{eq:rhos}
\end{align}
These nonperturbative parameters must be extracted from data.   
Neglecting $\rho^{sf}_1$ and $\rho^{sf}_8$, the combination $\rho^{sm}_1+\rho^{sm}_8/8$ was determined to be $0.15$ by fitting to the 
E687 and E691 fixed-target photoproduction data~\cite{Braaten:2001uu}. Neglecting $\rho^{sm}_8$, $\rho^{sf}_1$ and $\rho^{sf}_8$, the parameter
$\rho^{sm}_1$ was determined to be $0.06$ by fitting to data from the E791 experiment~\cite{Braaten:2002yt}. 
In this paper, we take $\rho^{sm}_1\sim0.06$
and $\rho^{sm}_8\sim0.7$. It turns out that these two contributions only account for $~10\%$ of the measured asymmetry 
$A_p=(-0.96\pm 0.26\pm0.18)\%$ at LHCb in Ref.~\cite{LHCb:2012fb}. Therefore, we include $\rho^{sf}_1$ and $\rho^{sf}_8$ and choose values of similar size as the spin-matched parameters. We use MSTW 2008 LO PDFs with $m_c = 1.275$ GeV and the Peterson parametrization for the fragmentation function~\cite{Peterson:1982ak} is used for $D_{c\rightarrow D^{\pm}}$:
\begin{equation}
D_{Q\rightarrow H}(z)=\frac{N_H}{z\left(1-\frac{1}{z}-\frac{\epsilon_Q}{1-z}\right)^2}.  \label{eq:Peterson}
\end{equation}
$\epsilon_c\sim (m_q/m_c)^2$ is chosen to be $0.06$. $N_H$ are determined by the averages of the measured
fragmentation probabilities listed in~\cite{Abramowicz:2013eja}. 
For the perturbative QCD rate, Eq.~(\ref{eq:pQCD}), which has no asymmetry if we ignore $C$ violation but enters into the denominator of     
Eq.~(\ref{eq:asymmetry}), we use the LO cross section and include feed down from $D^*$. The factorization scale is set to be $\mu_f=\sqrt{p_T^2+m_c^2}$.  

\begin{figure}
\begin{center}
\subfigure[]{
      \includegraphics[width=0.7\textwidth,angle=0,clip]{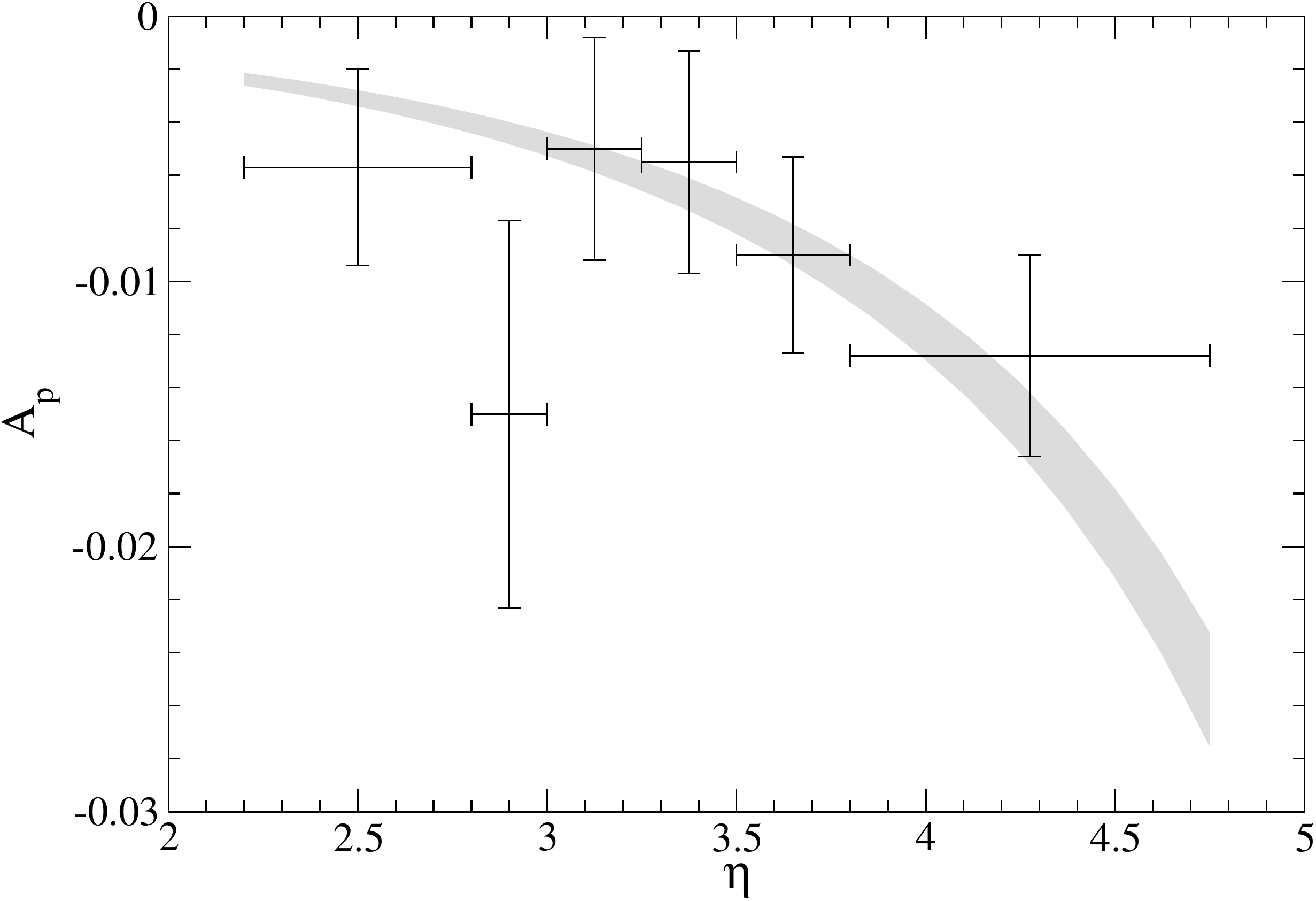}
}
\vskip .3in
\subfigure[]{
      \includegraphics[width=0.7\textwidth,angle=0,clip]{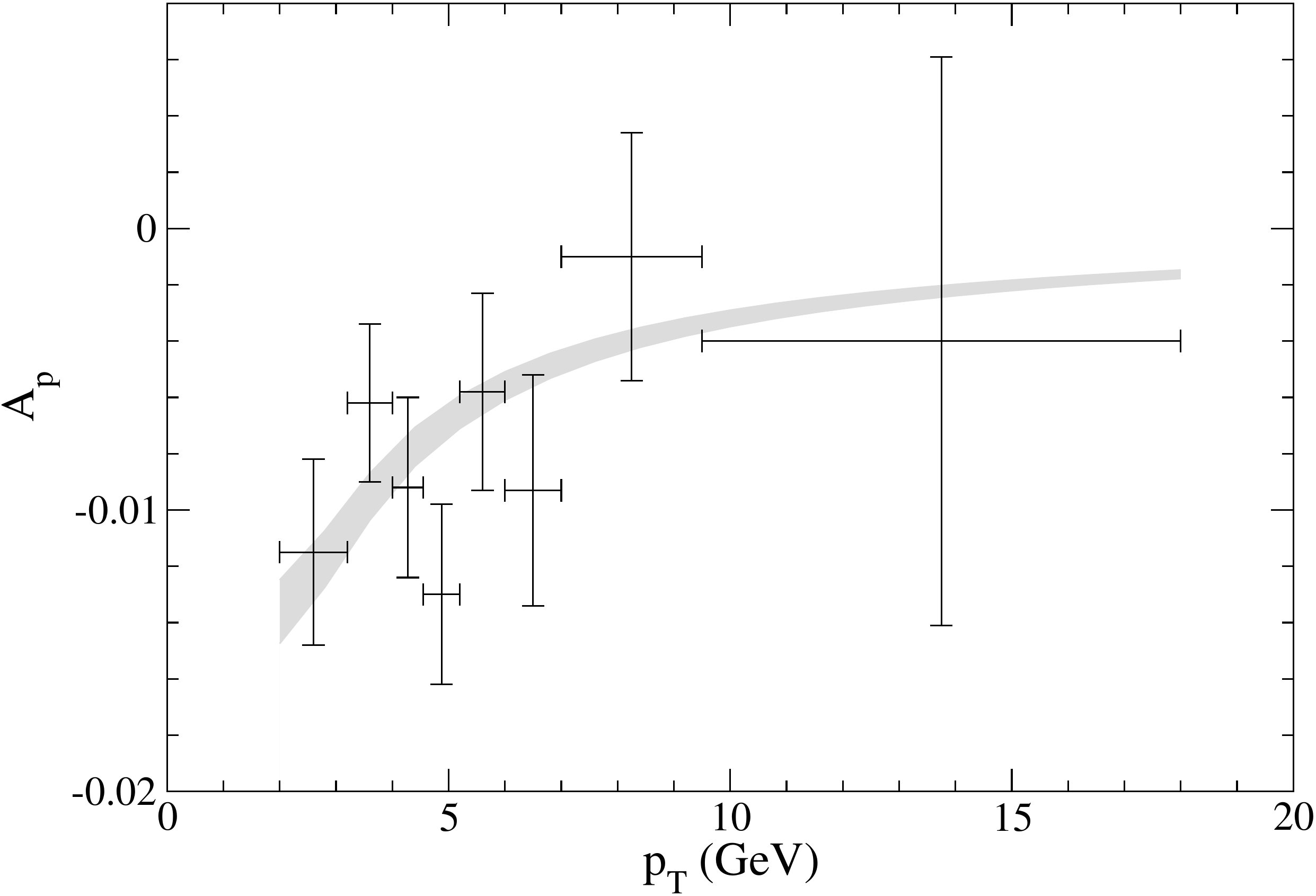}
}

\vskip .1in  
\caption[]{Asymmetry in $D^\pm$ production $A_p$ as a function of (a) pseudorapidity $\eta$ and (b) transverse momentum $p_T$ in $7$~TeV $pp$ collisions. The data points are from LHCb~\cite{LHCb:2012fb}. The grey band is obtained by varying the $\rho$s in the intervals
$0.055<\rho^{sm}_1<0.065$, $0.65<\rho^{sm}_8<0.8$, $0.4<\rho^{sf}_1<0.48$ and $0.4<\rho^{sm}_1<0.48$ respectively.}
\label{fg:Ap_dis}
\end{center}
\end{figure}

When integrated over $2\textrm{ GeV}<p_T<18\textrm{ GeV}$ and $2.2<\eta<4.75$, excluding the region with
$2\textrm{ GeV}<p_T<3.2\textrm{ GeV}$, $2.2<\eta<2.8$, the asymmetry $A_p$ for $D^\pm$ is found to be $-0.88\%<A_p<-1.04\%$ with $0.055<\rho^{sm}_1<0.065$, $0.65<\rho^{sm}_8<0.8$, $0.4<\rho^{sf}_1<0.48$ and $0.4<\rho^{sm}_1<0.48$. Figure~\ref{fg:Ap_dis} shows $A_p$
as a function of pseudorapidity $\eta$ and transverse momentum $p_T$ of the $D^\pm$ mesons as predicted by the heavy-quark recombination mechanism. Data from Ref.~\cite{LHCb:2012fb} are shown as well. The calculated distributions are reasonably consistent with the data.

In summary, we have calculated the $D^\pm$ asymmetry using the heavy-quark recombination mechanism for production at the LHCb experiment.  The measured asymmetry of $A_p=-0.96\pm0.26\pm0.18\%$ in the kinematic range $2.0\textrm{ GeV}<p_T<18\textrm{ GeV}$ and $2.2<\eta<4.75$ \cite{LHCb:2012fb} can be reproduced using reasonably sized non-perturbative parameters $\rho_{1,8}^{sm, sf}$.  Further, the $p_T$ and $\eta$ distributions are simultaneously reproduced by the heavy-quark recombination mechanism.

%===============================================================
% acknowledgements
%===============================================================
\section*{Acknowledgements}
We thank Tao Han and Thomas Mehen for useful discussions. AKL and WKL are supported in part by the National Science Foundation under Grant No. PHY-1212635.  AAP is supported in part by the U.S. Department of Energy under contract DE-FG02-12ER41825.

%===============================================================
% references
%===============================================================

\end{document}